\documentclass[preprint,12pt]{elsarticle}
\usepackage{amssymb}
\usepackage{amsmath}
\usepackage{algorithm}

\journal{Computer and Electrical Engineering}

\begin{document}

\begin{frontmatter}

\title{Delay Optimization of a Federated Learning-based UAV-aided IoT network}

\author{Hossein Mohammadi Firouzjaei} 
\affiliation{organization={University of Oulu},
            addressline={}, 
            city={Oulu},
            postcode={90570}, 
            state={North ostrobothnia},
            country={Finland}}

\author{Javad Zeraatkar Moghaddam} 
\affiliation{organization={University of Birjand},
            addressline={}, 
            city={Birjand},
            postcode={90000}, 
            state={Khorasan Jounobi},
            country={Iran}}

\author{Mehrdad Ardebilipour} 
\affiliation{organization={K. N. Toosi University of Technology},
            addressline={}, 
            city={Tehran},
            postcode={90500}, 
            state={Tehran},
            country={Iran}}            

\begin{abstract}
This paper explores the integration of power splitting(PS) simultaneous wireless information and power transfer (SWIPT) architecture and federated learning (FL) in Internet of Things (IoT) networks. The use of SWIPT allows power-constrained devices to simultaneously harvest energy and transmit data, addressing the energy limitations faced by IoT devices. The proposed scenario involves an Unmanned Arial Vehicle (UAV) serving as the base station (BS) and edge server, aggregating weight updates from IoT devices and unicasting aggregated updates to each device. The results demonstrate the feasibility of FL in IoT scenarios, ensuring communication efficiency without depleting device batteries.
\end{abstract}

\begin{graphicalabstract}
\begin{figure}[!t]
\setlength{\tabcolsep}{2pt}
\centerline{\includegraphics[width=4.5in]{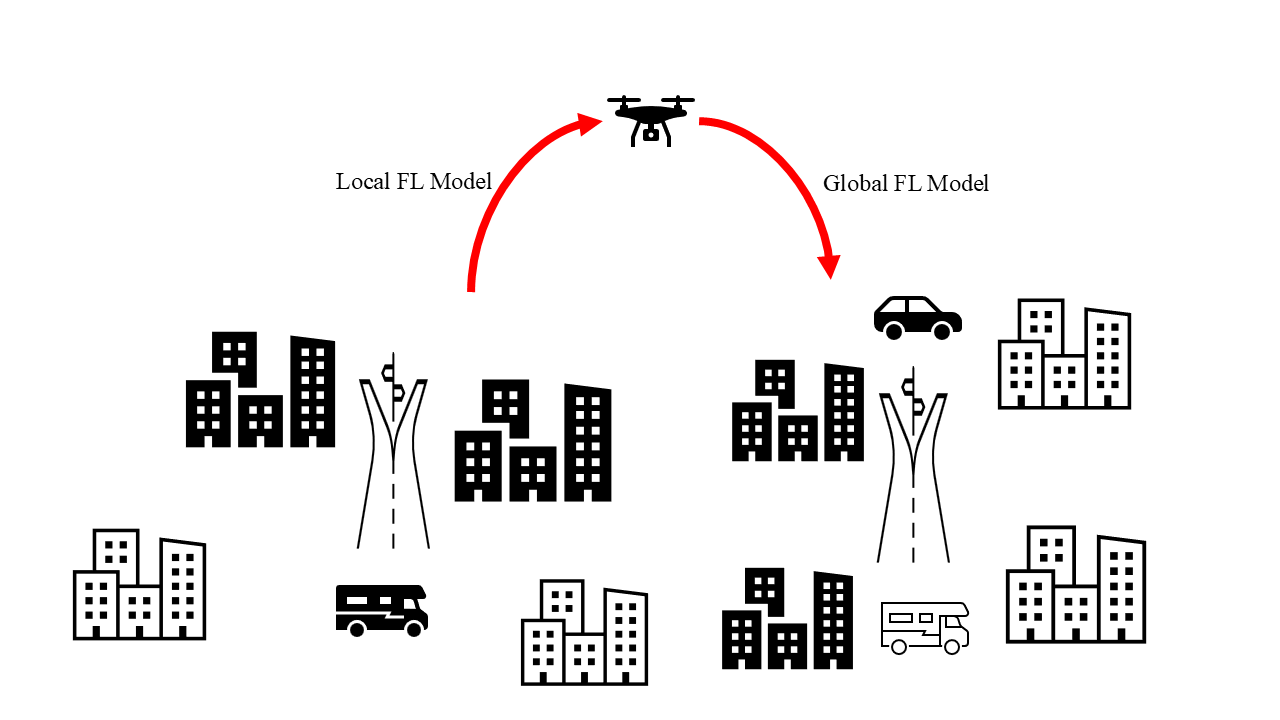}}
\caption{Graphical Abstract}
\label{Graphical Abstract}
\end{figure}
\end{graphicalabstract}

%%Research highlights
\begin{highlights}

\item Integration of SWIPT and Federated Learning for IoT Networks: This study presents a novel architecture integrating Power Splitting (PS) Simultaneous Wireless Information and Power Transfer (SWIPT) with Federated Learning (FL) in Internet of Things (IoT) networks. The approach allows IoT devices to harvest energy while transmitting data, effectively addressing the energy limitations of power-constrained devices, enhancing communication efficiency, and extending the operational lifespan of IoT systems.
\item UAV-Aided Model Aggregation and Energy Harvesting: The paper explores the use of Unmanned Aerial Vehicles (UAVs) as base stations and edge servers to aggregate model updates from IoT devices. The UAVs also broadcast energy to these devices, allowing for simultaneous data transmission and energy harvesting. The results show improved communication efficiency, energy harvesting, and device longevity, demonstrating the feasibility of Federated Learning in such energy-constrained environments.
\item Optimization Techniques for Reducing Communication Delay: The study develops learning-based methods to optimize communication delays in Federated Learning-based IoT networks. By utilizing UAVs for model aggregation and communication, the proposed model significantly reduces delay while maintaining efficient energy consumption, enabling more scalable and sustainable IoT systems.

\end{highlights}

\begin{keyword}
UAV \sep SWIPT \sep IoT \sep FL
\end{keyword}

\end{frontmatter}

\section{Introduction}
The rapid growth of IoT devices has led to a vast amount of data with potential applications across various domains [1], [2]. However, integrating machine learning into IoT devices faces challenges due to distributed data and communication constraints [3], [4]. Federated Learning (FL) has emerged as a solution prioritizing privacy, data heterogeneity, and communication efficiency. Energy limitations in IoT devices remain an obstacle [5]. Energy harvesting (EH), like PS Wireless Power Transfer (PS-WPT or SWIPT), offers a promising approach. Still, ensuring sufficient energy for FL training needs further exploration [6], [7].

SWIPT enables EH and data transmission in wireless systems, benefiting FL [5]. FL involves devices collaboratively training a global model while preserving data locally. Incorporating SWIPT allows power-constrained devices to harvest energy while transmitting model updates, addressing energy challenges in FL [4]. It extends the operational lifespan of power-constrained devices, like battery-powered IoT devices, by replenishing energy during communication. PS SWIPT in FL tackles energy constraints, enhances device lifespan, improves communication reliability, increases energy efficiency, enables scalability, and promotes sustainability [7]. SWIPT empowers devices to harvest energy while transmitting data, facilitating FL implementation in power-constrained environments.

This paper investigates the performance of an IoT network implemented using the PS SWIPT architecture and FL design. In the proposed scenario, a UAV serves as the base station (BS) or edge server, responsible for transmitting aggregated weights and energy. Aggregating the updated weight from devices and unicasting these updates to the devices are the main roles of this edge server. Meanwhile, each IoT device harvests energy from the received signal to perform its local computations. The results demonstrate the feasibility of FL in IoT scenarios, showcasing improved communication efficiency without depleting device batteries.
\subsection{Related Work}
The authors of [1] have presented a novel policy for joint client scheduling and resource block (RB) allocation in FL over wireless links with imperfect channel state information (CSI). The proposed policy utilizes a Gaussian process regression (GPR)-based channel prediction method and employs the Lyapunov optimization framework to minimize accuracy loss in FL. Numerical simulations demonstrate up to a $25.8\%$ reduction in accuracy loss compared to existing client scheduling and RB allocation methods. This work contributes to addressing the challenges of FL under imperfect CSI and provides an effective solution for optimizing client scheduling and RB allocation. Future extensions could involve analyzing computation-communication tradeoffs and investigating the impact of communication errors. Overall, the proposed approach enhances the performance of FL in wireless scenarios and advances distributed learning techniques.

Minsu Kim et al. [7] have addressed the challenge of deploying FL over resource-constrained wireless networks, where balancing accuracy, energy efficiency, and precision is crucial. To overcome the impracticality of using high-precision deep neural networks (DNNs) on such devices, a quantized FL framework is proposed. The framework employs quantized neural networks (QNNs) to represent data with finite precision during local training and uplink transmission. Rigorous energy models are derived for both local training and transmission, and an optimization problem is formulated to minimize energy consumption while ensuring convergence. The paper analytically derives the FL convergence rate and employs a line search method for solution finding.

Zhaohui Yang et al. [3] have addressed the problem of energy-efficient transmission and computation resource allocation for FL over wireless communication networks. The model considers users with limited local computational resources training their FL models and transmitting them to a base station (BS) for aggregation. The goal is to minimize total energy consumption while meeting latency constraints. An iterative algorithm is proposed, providing closed-form solutions for time allocation, bandwidth allocation, power control, computation frequency, and learning accuracy. To obtain an initial feasible solution, a bisection-based algorithm is used. Numerical results demonstrate up to a $59.5\%$ reduction in energy compared to conventional FL methods. Simulation results highlight the scheme's superiority, particularly for low maximum average transmit power.

Authors in [8] have explored the deployment of FL in an EH wireless network with a base station (BS) employing massive MIMO. They address the joint energy management and user scheduling problem to optimize FL training loss while considering interference and energy constraints. The convergence rate of the FL algorithm is analyzed to understand the impact of factors like transmit power and scheduled users on training loss. The proposed user scheduling and energy management schemes are shown to reduce training loss compared to standard FL algorithms. The study also extends to multiple BSs, addressing the joint user association and scheduling problem using the branch-and-bound technique. Simulation results demonstrate the effectiveness of the proposed resource management schemes in EH wireless systems for FL. Overall, this work presents a novel EH FL framework, contributing to the understanding and optimization of FL in wireless networks with EH capabilities.
\subsection{Motivations and Key Contributions}
Although such efforts are done, several challenges still exist in current works, which are given as follows:
\begin{itemize}
  \item How can FL be implemented in a UAV-enabled cognitive radio network with EH capabilities?
  \item What are the key performance metrics that should be considered for the proposed system?
  \item How do different EH techniques and parameters affect the performance of the proposed system?
\end{itemize}

Motivated by the aforementioned challenges, a novel SWIPT-based FL model for UAV-aided IoT networks is proposed in this work. In summary, the significant contributions of this work are as follows:
\begin{enumerate}
  \item Developing a mathematical model for the EH implemented FL-based UAV-enabled IoT network.
  \item Developing learning-based methods for optimizing the delay of FL-based IoT network.
  \item Investigating the optimization at the local devices/UAVs and evaluating the performance of the proposed methods using simulations.
\end{enumerate}

\subsection{Organization}
The system model is investigated In section II, which comprises the communication model, EH model, and FL model. Additionally, the convergence analysis of the FL model training process is defined in section II. After that, an algorithm is in proposed section III to optimize the total delay. Experimental Analysis and Results are studied in section IV. Finally, our work is concluded in section V.
\section{System Model}
\begin{figure}[!t]
\setlength{\tabcolsep}{2pt}
\centerline{\includegraphics[width=2.5in]{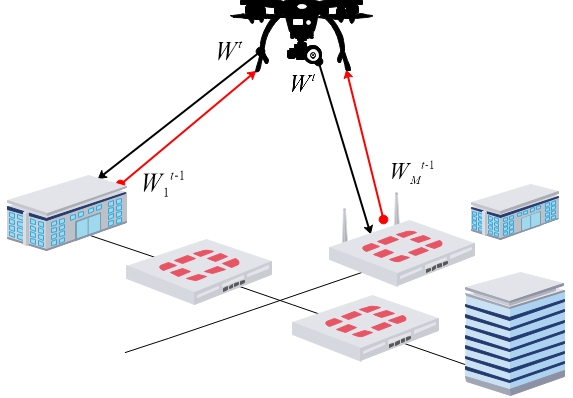}}
\caption{System Model}
\label{System model}
\end{figure}

The system model of this paper is shown in Fig.\ref{System model} and includes an EH model, uplink and downlink telecommunication channel model, and machine learning model. The communication channel model includes uplink and downlink channels, which by defining the channel coefficients and Riley's feeding channel model, the value of the communication rate and the total time of the communication obtained. In the EH model, considering the PS architecture for the receiver, the amount of energy harvested and the duration of communication are calculated.

The communication model, FL approach, and PS-SWIPT design of the proposed scenario are studied in the following sections to evaluate the performance of an FL-based PS-SWIPT UAV-aided IoT network.

\subsection{Communication Model}
The communication model describes the interaction between the UAV and IoT devices. In the uplink communication model, users establish a connection with the UAV using the Rayleigh fading channel. In this model, we assume that the transfer power of each user in the uplink, denoted as $P_t^u$, remains constant for all users. The received signal power by the UAV from the $i_{th}$ user, $P_{r,i}^u$, can be mathematically expressed as follows:
\begin{equation}
  P_{r,i}^u =  P_t^u R_{i,u}^{-\alpha _u} |g_i|^2,
\end{equation}
where $R_{i,u}$ denotes the distance between the $i_{th}$ user and the UAV, $\alpha _u$ indicates the path-loss exponent and $g_i \ \epsilon \ C^{H\times1}$ is the complex channel vector between the $i_{th}$ user and the UAV.

Then, the power of the interference of other users for the uplink signal of $i_{th}$ user, $I_{i}^u$, can be calculated as follows:
\begin{equation}
  I_{i}^u =  \sum_{j,\ j \ne i}^{M}P_t^u R_{j,u}^{-\alpha _u} |g_j|^2,
\end{equation}
where $R_{j,u}$ denotes the distance between the $j_{th}$ user and the UAV, $g_j \ \epsilon \ C^{H\times1}$ is the complex channel vector between the $j_{th}$ user the UAV and $M$ denotes the number of IOT devices.

Hence, The signal-to-interference-plus-noise ratio (SINR) of the $i_{th}$ uplink user at the UAV can be indicated as:
\begin{equation}
  \gamma_{i}^u = \frac{P_{r,i}^u}{I_{i}^u + N^u},
\end{equation}
where $N^u$ is the noise power at the UAV in uplink communication.

After that, the achievable rate of the device the $i_{th}$ uplink user at the UAV is given as follows:
\begin{equation}
  \zeta _i^u = W log_2(1+\gamma _i^u),
\end{equation}
where $W$ indicates the transmission bandwidth.

Finally, the time length of the uplink communication between the $i_{th}$ user at the UAV can be calculated as follows:
\begin{equation}
  T _{i}^u = \frac{Data^u}{\zeta_i^u},
\end{equation}
where it is assumed that each user has $Data^u$ bits to transmit in uplink communication.

On the other hand, the FL approach and SWIPT techniques are established in the downlink. The received power at the $i_{th}$ user can be expressed as follows:
\begin{equation}
  P_{r,i}^d =  P_t^d R_{i,u}^{-\alpha _u} |g_i|^2,
\end{equation}
where $P_t^d$ is the transfer power of the UAV.

$\delta_i$ indicates the PS parameter of the $i_{th}$ user ($\delta_i \epsilon (0,1)$).

$\delta_i P_{r,i}^d$ denotes the amount of power that dedicates to the data decoding section. On the other hand, $(1- \delta_i) P_{r,i}^d$ is dedicated to SWIPT section.

Then, the SINR at the $i_{th}$ user can be calculated as:
\begin{equation}
  \gamma_{i}^d = \frac{\delta_i P_{r,i}^d}{I_{i}^d + N^d},
\end{equation}
where $N^d$ is the power of WAGN at the UAV in downlink communication and $I_{i}^d$ is the power of the interference of other users for the downlink signal of the UAV that is received at the $i_{th}$ user, which can be expressed as follows:
\begin{equation}
  I_{i}^d =  \sum_{j,\ j \ne i}^{M}P_t^d R_{j,u}^{-\alpha _u} |g_j|^2,
\end{equation}

Following that, the achievable rate at the $i_{th}$ user in downlink can be expressed as:
\begin{equation}
  \zeta _i^d = W log_2(1+\gamma _i^d),
\end{equation}

The time length of the downlink communication between the UAV and the $i_{th}$ user can be calculated as follows:
\begin{equation}
  T _{i}^d = \frac{Data^d}{\zeta_i^d},
\end{equation}
where it is assumed that the UAV has $Data^d$ bits to transmit in downlink communication.

\subsection{FL Model}
The UAV equipped with a mobile edge computing (MEC) server provides computing services and broadcasts energy to the users.  Each user has an onboard computing processor to train the local model using its data and has an energy circuit to harvest energy transmitted by the UAV. To leverage the distributed nature of the devices and preserve data privacy, we employ FL as our collaborative training approach. The FL model allows devices to perform local model training using their respective data and share only the updated model weights with the UAV. The UAV aggregates these weights to create a global model that captures the collective intelligence of all devices. This FL model ensures efficient knowledge transfer while maintaining data privacy and reducing the communication overhead between the UAV and devices.

$W$ is considered as the model parameter for the global model of the edge server. Moreover, $W_i$ indicates the parameters of the local model for the $i_{th}$ user. In addition, $D_i$ denotes the set of training datasets that are used by the $i_{th}$ user. Hence, the loss function $f(W; q_{i,j}, v_{i,j})$ is defined to measure the error in FL performance across the input data sample vector $q_{i,j}$ regarding the learning model $w$ and the target output value $v_{i,j}$ for every $j_{th}$ input sample when considering the $i_{th}$ user. The sum loss function for the $i_{th}$ user on $D_i$ is defined as follows:
\begin{equation}
  F_i(W) = \frac{1}{|D_i|} \sum_{j \epsilon D_i} f(W; q_{i,j}, v_{i,j}), \forall i\ (i = 1, 2, ..., N),
\end{equation}
Where $N$ indicates the number of users. The average global cost function utilizing decentralized individual datasets at the UAV server can be expressed as:
\begin{equation}
  F(W) \triangleq \frac{1}{\sum_{i}|D_i|} \sum_{i=1}^{N} \sum_{j \epsilon D_i} f(W; q_{i,j}, v_{i,j})
\end{equation}

The objective of the FL task is to search for the optimal model parameters at the UAV server that minimizes the global loss function as:
\begin{equation}
  W^* = argmin \ F(W)
\end{equation}

$F(W)$ is not readily calculable using the original datasets from chosen devices, as this would jeopardize the privacy and security of each device. Therefore, to tackle the problem in (13), a distributed approach can be employed. In the following discussion, we focus on the implementation of FL model-averaging, although the same principle can be applied to achieve an alternative implementation using the gradient-averaging method.

The predefined stopping criterion for the FL model is crucial to ensure that the training process stops at an appropriate point, preventing overfitting and unnecessary communication overhead. A common stopping criterion in FL is based on the number of communication rounds or epochs. The FL training process stops after a predefined number of communication rounds ($R^c$) or epochs.

During each communication round, the UAV communicates with the IoT devices, collects their local model updates, and aggregates them to form a new global model. After every communication round, the global model is redistributed to the IoT devices, and they start the next round of local training. By setting a predefined number of communication rounds ($R^c$), the FL process terminates after a certain number of iterations. The value of $R^c$ should be determined based on the convergence rate of the FL process and the resource constraints of the network. It can be set through cross-validation on a validation dataset.

By employing this predefined stopping criterion, the FL process will automatically terminate after $R^c$ communication rounds, and the UAV will have a trained global model that captures the collective intelligence of all devices while preserving data privacy and minimizing communication overhead. To set the predefined number of communication rounds ($R^c$) based on cross-validation on a validation dataset, we have considered the following steps:
\begin{enumerate}
  \item Data Preparation: Split the dataset into three parts: training data for local training ($D_{\text{train}}$), validation data for cross-validation ($D_{\text{val}}$), and test data for final evaluation ($D_{\text{test}}$).
  \item Initialize Parameters: Decide on a range of possible values for $R^c$ that you want to explore, such as [5, 10, 15, 20, ...].
  \item Cross-validation Loop: For each value of $R^c$ in the range, perform the following steps:
            \begin{itemize}
              \item FL Training: Run the FL training process with $R^c$ communication rounds, aggregating the local model updates, and redistributing the global model.
              \item Validation: Evaluate the performance of the global model on the validation dataset ($D_{\text{val}}$). Calculate relevant evaluation metrics, such as accuracy, loss, etc.
              \item Record Metrics: Keep track of the performance metrics obtained for each value of $R^c$.
            \end{itemize}
  \item Analyze Results: Examine the performance metrics recorded during the cross-validation loop. Look for trends in the performance as $R^c$ increases.
  \item Choose Optimal $R^c$: Select the value of $R^c$ that provides the best performance on the validation dataset. This may be the point where the performance stabilizes or plateaus.
  \item Evaluate on Test Dataset: Use the final global model obtained with the selected $R^c$ value to evaluate its performance on the test dataset ($D_{\text{test}}$). This gives an unbiased estimate of the model's generalization ability.
\end{enumerate}

By using cross-validation on the validation dataset, we could effectively select the optimal number of communication rounds ($R^c$) for the FL process. This approach helps to avoid overfitting the training data and ensures that the selected value of $R^c$ generalizes well to unseen data. Note that we have performed multiple cross-validation runs to ensure the stability of our results. The optimal $R^c$ may vary based on the specific characteristics of the dataset and the complexity of the problem that is addressed in the UAV-aided IoT network.

\subsection{EH Model and Training Model Update}
EH plays a vital role in prolonging the UAV's flight time. We incorporate an EH model into our system to enable the UAV to harvest energy from the environment. Specifically, we explore the concept of SWIPT to efficiently transfer data and harvest energy simultaneously.

The energy consumption to compute the model at the $i_{th}$ user can be given as follows:
\begin{equation}
  E_i^c = i C_i A_i I_i f_i^2,
\end{equation}
where the Efficient switched capacitance of the $i_{th}$ user is named as $i$, $C_i$ denotes the computational workload for processing a single data sample in terms of central processing unit cycles in $Mcycles/bit$ ,$A_i$ indicates the data size of the $i_{th}$ user measured in bits, $I_i$ is the training algorithm's local iteration count and $f_i$ denotes the clock speed of the processor of the $i_{th}$ user in GHz.

The energy to transmit the weights in the uplink communications can be given as follows:
\begin{equation}
  E_i^u = T _{i}^u P_t^u
\end{equation}

In addition, the energy that the UAV spends to communicate to the $i_{th}$ user can be calculated as:
\begin{equation}
  E_i^d = T _{i}^d P_t^d
\end{equation}

Hence, the consumed energy by the $i_{th}$ user can be defined as follows:
\begin{equation}
  E_i^{total} = E_i^c + E_i^u + E_i^d
\end{equation}

The harvested power in the $i_{th}$ user during one communication round can be defined by a nonlinear EH model as follows:
\begin{equation}
  P_i^H = \alpha _1 ((1 - \delta _i) \times P_{r,i}^d)^2 + \alpha _2 ((1 - \delta _i) \times P_{r,i}^d) + \alpha _3 ,
\end{equation}
where $\alpha _1 ,\alpha _2, \alpha _3 \epsilon R$ are the parameters of the model.

We define the energy harvested at the $i_{th}$ user as $E_i^H = T _{i}^d P_i^H$, where $T _{i}^d$ is the transmission time in seconds that the edge server spends to unicast the aggregated weight to device k.

Hence, the total energy spent at $i_{th}$ user during one communication round should be lower than The harvested energy in the $i_{th}$ user during one communication round: $E_i^{total} \le E_i^H$

In addition, the duration required for local training task completion at the $i_{th}$ user can be given as:
\begin{equation}
  T_i^L = C_i A_i I_i/f_i
\end{equation}

On the other hand, the UAV server takes $T_{UAV} = C_u Data^u/f^u$ for the global model aggregation, where $C_u$ represents the count of CPU cycles utilized during model training on a single data sample at the UAV server, while $f^u$ represents the computation capability (CPU cycles per second) of the UAV server."

At last, the period for a single communication round, encompassing the overall time expenditure for organizing the FL model, uplink transmission with local computation, and downlink transmission with harvesting time, can be expressed as follows:
\begin{equation}
  t^{total}(r) = max_i(T _{i}^u + T_i^L) + max_k(T _{i}^d) + T_{UAV}
\end{equation}

\section{Proposed FL-based SWIPT aided Algorithm}
Minimizing the total delay of communication in an FL-based UAV-aided IoT network involves several key components. The main parts of achieving this goal can be mentioned UAV Deployment and Mobility Management, IoT Device Clustering, Local Model Training, FL Protocol, Quality of Service (QoS) Management, and Efficient Model Aggregation. By addressing these main parts, the total delay of communication in a FL-based UAV-aided IoT network can be minimized, leading to more efficient and effective collaboration between the UAV and IoT devices for machine learning tasks.

To offer novelty for this paper that employs SWIPT and FL methods, the following research directions and ideas are considered:
\begin{itemize}
  \item Privacy-Preserving SWIPT-FL: Address privacy concerns in SWIPT-FL systems by developing privacy-preserving FL algorithms that allow devices to securely share their model updates while minimizing the disclosure of sensitive information.
  \item SWIPT for Model Aggregation: Investigate the use of SWIPT for model aggregation in FL. Explore how devices can contribute to the aggregation process by sending model updates while simultaneously harvesting energy, reducing the need for central servers with continuous power supply.
\end{itemize}

As studied in the previous section, Let's integrate the predefined stopping criterion into the 5-step algorithm for minimizing the total delay of communication in a FLbased UAV-aided IoT network. The proposed algorithm with one UAV and $M$ IoT users is given in algorithm $1$.
\begin{algorithm}

    \begin{tabular}{p{230pt}}
        \textbf{Algorithm1: FL-based SWIPT aided Algorithm}\\
        \hline
    \end{tabular}
    \\
    \textbf{Step 1}: UAV Deployment and Initialization
    \\  \ \ \ Date1: Setting the UAV Location
    \\  \ \ \ Data2: computing $R_{i,u}$ for each Device
    \\  \ \ \ Data3: Initializing the global machine learning model $W$ at the UAV
    \\
    \textbf{Step 2}: Model Initialization (IoT Device Clustering)
    \\  \ \ \ Data1: Selecting $M$ IoT Devices
    \\
    \textbf{Step 3}: Local Model Training and FL
    \\  \ \ \ UAV sends the global model $W$ to all IoT devices
    \\  \ \ \ Each IoT device $i$ performs local model training using its data $D_i$.
    \\  \ \ \ Each IoT device $i$ performs SWIPT techniques
    \\  \ \ \ IoT device $i$ calculates the local model weights $W_i$.
    \\  \ \ \ IoT device $i$ sends $W_i$ back to the UAV.
    \\  \ \ \ UAV aggregates the received weights to update the global model $W$ (using Federated Averaging).
    \\
    \textbf{Step 4}: Model Update and Communication
    \\  \ \ \ communicating the final global model $W$ to all $M$ IoT users
    \\
    \textbf{Step 5}: IoT Device Inference and Decision Making
    \\  \ \ \ Using the received global model $W$ for making inferences on local data
    \\  \ \ \ Utilizing the updated global model for decision making
    {
    \\
    Repeat Steps 3 to 5 for multiple rounds of training until the model converges or reaches a predefined stopping criterion
    }
\end{algorithm}

In step $1$, as expected and discussed in our previous work [6], the strategic location for optimal coverage and communication range to all $M$ IoT users is the center of the area. In addition, Initializing the global machine learning model $W$ at the UAV is done in this step (with random weights). Following that and in Step 2, the $M$ IoT devices into clusters based on their proximity to the UAV or their communication capabilities are selected. This step tries to ensure that each UAV is assigned to manage a manageable number of IoT devices for efficient communication.

Then, Local Model Training and FL are defined in Step 3. Performing iterative FL rounds with a predefined stopping criterion $R^c$ based on cross-validation on a validation dataset is designed as shown in step 3 of the algorithm. After the predefined number of communication rounds ($R^c$) or convergence, the FL process has to be stopped.

Model Updates and Communication are given in Step 4. The UAV communicates the final global model $W$ to all $M$ IoT users. IoT Device Inference and Decision-Making are shown in Step 5. Each IoT device uses the received global model $W$ for making inferences on its local data. Utilizing the updated global model for decision making belongs to this step.

By incorporating the predefined stopping criterion based on cross-validation on the validation dataset, the algorithm ensures that the FL process terminates after a suitable number of communication rounds. This approach helps prevent overfitting and minimizes the total communication delay in the FL-based UAV-aided IoT network while still achieving a well-generalized global model. The value of $R^c$ should be determined through experimentation and analysis, considering the dataset characteristics, problem complexity, and available resources.

\section{Simulation Results}
The simulation results of this paper are presented in this section to evaluate the impact of using SWIPT techniques and FL algorithm on the performance of an IoT system. In fact, in this section, a comparison is made between the scenario proposed in this paper and the existing scenarios that do not use machine learning and EH as done in this paper.

\begin{figure}[!t]
\setlength{\tabcolsep}{4pt}
\centerline{\includegraphics[width=4.5in]{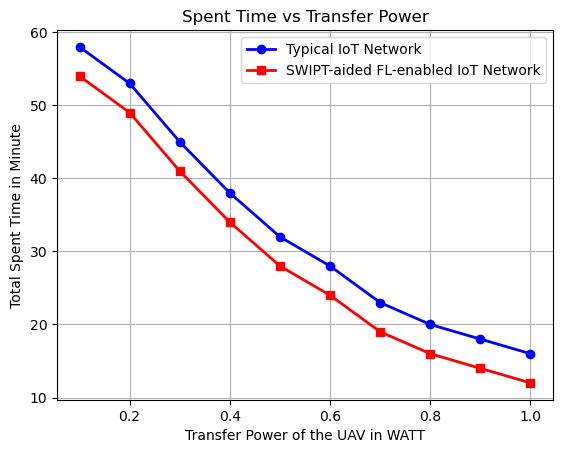}}
\caption{the amount of energy consumed versus the transmission power of users}
\label{Simulation figure 1}
\end{figure}

Fig.\ref{Simulation figure 1} shows the relationship between the duration of a communication round and the UAV's transmission power. As it was predicted from the theoretical results, with the increase in the power of sending the UAV, the time duration decreases. There are two reasons for this inverse relationship between the connection time and the UAV transmission power. The first reason is that with the increase in the UAV transmission power, the communication rate between the UAV and IoT users increases, and the weight updates are collected in a shorter period. The second reason is that by increasing the transmission power of the UAV, IoT users can harvest more energy in a shorter period of time and perform local calculations faster. In addition, the effective impact of the scenario proposed in this paper is observed to enhance the efficiency of IoT networks. Figure 2 shows that the scenario proposed in this paper performs better than the usual scenario.

\begin{figure}[!t]
\setlength{\tabcolsep}{4pt}
\centerline{\includegraphics[width=4.5in]{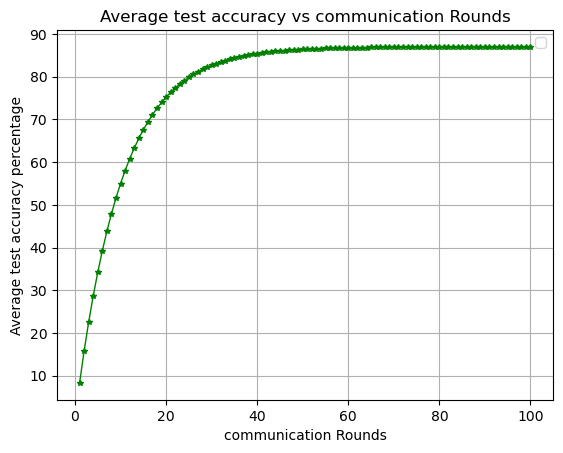}}
\caption{the amount of energy consumed versus the transmission power of users}
\label{Simulation figure 2}
\end{figure}

Fig.\ref{Simulation figure 2} illustrates the average test accuracy plotted against the number of communication rounds across 1000 Monte Carlo iterations. The figure depicts the relationship between the average test accuracy and the number of communication rounds, confirming the alignment with theoretical predictions. As the number of communication rounds increases, there is a noticeable rise in the average test accuracy. This finding validates the expected outcomes based on theoretical analysis. The upward trend indicates that increased communication rounds positively impact the accuracy of the test results. The graph demonstrates a consistent and favorable pattern, bolstering confidence in the validity of the theoretical conclusions. Overall, the figure supports the hypothesis that more communication rounds lead to improved average test accuracy, reinforcing the theoretical basis of the study.

\section{Conclusion}
Incorporating PS SWIPT architecture into FL offers significant advantages for IoT networks. By enabling devices to harvest energy while transmitting data, the proposed approach overcomes energy limitations, extends device lifespan, enhances communication reliability, improves energy efficiency, and promotes sustainability. The feasibility of FL in IoT scenarios is demonstrated through the implemented UAV-based system, highlighting the potential for efficient and long-lasting IoT deployments. With further advancements in PS SWIPT and FL techniques, the vision of energy-aware and scalable IoT networks can be realized.

\end{document}